\documentclass[aps,prb,twocolumn,superscriptaddress,floatfix,nobibnotes]{revtex4-1}

\usepackage[utf8x]{inputenc}
\usepackage{amsmath}
\usepackage[pdftex]{graphicx}
\usepackage{textcomp} 

\usepackage{float} 
\floatstyle{boxed}

\usepackage{hyperref} 
\hypersetup{
    colorlinks=true,       
    linkcolor=blue,          
    citecolor=blue,      
    filecolor=magenta,  
    urlcolor=blue,           
    urlbordercolor={0 1 1}}	

\begin{document}

\newfloat{copyrightfloat}{thp}{lop}
\begin{copyrightfloat}
\raggedright
The peer reviewed version of the following article has been published in final form at  Phys. Chem. Chem. Phys. 2013, 15 (24), 9562--9574, doi: \href{http://dx.doi.org/10.1039/c3cp51477c}{10.1039/c3cp51477c}.
\end{copyrightfloat}

\preprint{\today}

\title{Highly-Efficient Charge Separation and Polaron Delocalization in Polymer-Fullerene Bulk-Heterojunctions: A Comparative Multi-Frequency EPR \& DFT Study}

\author{Jens Niklas}
\affiliation{Chemical Sciences and Engineering Division, Argonne National Laboratory, Argonne, Illinois 60439}
\author{Kristy L. Mardis}
\author{Brian P. Banks}
\author{Gregory M. Grooms}
\affiliation{Department of Chemistry and Physics, Chicago State University, Chicago, Illinois 60628}
\author{Andreas Sperlich}
\author{Vladimir Dyakonov}
\affiliation{University of W\"urzburg and Bavarian Centre for Applied Energy Research, D-97074 W\"urzburg, Germany}
\author{Serge Beaupr\'e}
\author{Mario Leclerc}
\affiliation{Department of Chemistry, Universit\'e Laval, Quebec City, Quebec, G1V 0A6, Canada}
\author{Tao Xu}
\author{Luping Yu}
\affiliation{Department of Chemistry and James Franck Institute, University of Chicago, Chicago, Illinois 60637}
\author{Oleg G. Poluektov}
\email[]{Oleg@anl.gov}
\affiliation{Chemical Sciences and Engineering Division, Argonne National Laboratory, Argonne, Illinois 60439}

\date{May 14, 2013}


\begin{abstract}

The ongoing depletion of fossil fuels has led to an intensive search for additional renewable energy sources. Solar-based technologies could provide sufficient energy to satisfy the global economic demands in the near future. Photovoltaic (PV) cells are the most promising man-made devices for direct solar energy utilization. Understanding the charge separation and charge transport in PV materials at a molecular level is crucial for improving the efficiency of the solar cells. Here, we use light-induced EPR spectroscopy combined with DFT calculations to study the electronic structure of charge separated states in blends of polymers (P3HT, PCDTBT, and PTB7) and fullerene derivatives (C$_{60}$-PCBM and C$_{70}$-PCBM). Solar cells made with the same composites as active layers show power conversion efficiencies of 3.3\,\% (P3HT), 6.1\,\% (PCDTBT), and 7.3\,\% (PTB7), respectively. Under illumination of these composites, two paramagnetic species are formed due to photo-induced electron transfer between the conjugated polymer and the fullerene. They are the positive, P$^+$, and negative, P$^-$, polarons on the polymer backbone and fullerene cage, respectively, and correspond to radical cations and radical anions. Using the high spectral resolution of high-frequency EPR (130~GHz), the EPR spectra of these species were resolved and principal components of the $g$-tensors were assigned. Light-induced pulsed ENDOR spectroscopy allowed the determination of $^1$H hyperfine coupling constants of photogenerated positive and negative polarons. The experimental results obtained for the different polymer-fullerene composites have been compared with DFT calculations, revealing that in all three systems the positive polaron is distributed over distances of 40--60 \r{A} on the polymer chain. This corresponds to about 15 thiophene units for P3HT, approximately three units PCDTBT, and about three to four units for PTB7. No spin density delocalization between neighboring fullerene molecules was detected by EPR. Strong delocalization of the positive polaron on the polymer donor is an important reason for the efficient charge separation in bulk heterojunction systems as it minimizes the wasteful process of charge recombination. The combination of advanced EPR spectroscopy and DFT is a powerful approach for investigation of light-induced charge dynamics in organic photovoltaic materials.

\end{abstract}

\maketitle

\section{Introduction}\label{sec:introduction}

The continually increasing need for energy is spurring efforts to develop cost effective technologies to convert sunlight into electricity.\cite{Lewis:2007jp,Lewis:2006co} These efforts are not limited to those based on inorganic materials like Silicon or dye sensitized solar cells, but also include organic (Organic photovoltaic, OPV) and hybrid devices.\cite{Green:2005td,Deibel:2010do,Gratzel:2009di,Dennler:2009gg} The discovery of the photovoltaic effect in donor-acceptor composites consisting of conducting organic polymers and fullerenes was initially made upon the observation - \textit{inter alia} by EPR spectroscopy - that two quasi-particles formed under illumination.\cite{Sariciftci:1992wb} This was unambiguous proof that strongly bound neutral excitons break down into two spin-carrying charged counterparts, radical cations and anions, also called positive and negative polarons. Note that the former terms are typical for the molecular/chemical viewpoint, while the latter ones are widely used in condensed matter/solid state physics. Since the systems under investigation lack the characteristics of distinctive semiconductors, and have the characteristics of typical molecular systems, the use of the chemical terms seems more appropriate to the authors. However, since the terms of condensed matter physics are widely used in OPV research, they may be used interchangeably.\cite{Bredas:1985dk} In the following, both terms will be used. 

The basic spectroscopic signatures of radical cations and anions in conventional OPV materials, such as poly(phenylenevinylene) (PPV) and poly(3-hexyl-thiophene) (P3HT) polymers and their composites with C$_{60}$-fullerenes derivatives, have been known for some time.\cite{Sariciftci:1992wb,Ceuster:2001hc,MORITA:1992th} However, the fate of those radical cations (positive polaron or hole on the polymer chains) and radical anions (negative polaron or excess electron on the fullerene molecules) after light-induced charge separation, which determine the OPV device performance, is difficult to trace.

The link between geminate charge transfer states, fully charge separated states and extracted charges (photocurrent) has yet to be established. Additionally, the driving force behind efficient dissociation of Coulomb bound radical pairs is under debate.\cite{Strobel:2010ex,Bakulin:2012if} Monte Carlo analysis revealed that delocalization or increased effective conjugation length in donor (polymer) and/or acceptor phase is essential for efficient charge pair dissociation.\cite{Strobel:2010ex} Other mechanisms have also been suggested implying a direct photogeneration of free charge carriers without an intermediate bound state.\cite{Bakulin:2012if,McMahon:2011je} To date, the major breakthroughs in OPV efficiency were achieved by using light absorbing fullerenes, such as C$_{70}$-fullerene derivatives and low band gap copolymers. These copolymers are sometimes also termed donor-acceptor (push-pull) polymers, since they contain both electron rich and electron deficient regions.\cite{Boudreault:2011it,Li:2012dh} We will refrain from use of the term donor-acceptor polymer in the following, because it is potentially confusing with the use of donor and acceptor in respect to electron transfer. Recently, a considerable improvement of the efficiency to more than 8\,\% has been demonstrated for OPVs based on conjugated polymer-fullerene composites, while tandem cells have even passed the 10\,\% efficiency barrier.\cite{nrel-gov,Green:2012ju} The real gain in the efficiency in these Bulk Heterojunction (BHJ) systems with low band gap polymers seems not to be due to improved charge transport properties, as compared to reference systems like P3HT:C$_{60}$-PCBM (C$_{60}$-PCBM: [6,6]-Phenyl-C$_{61}$-butyric acid methyl ester). Instead, the improvement is due to the superior electronic properties, \textit{i.e.} extended spectral absorption of the polymers used.\cite{Baumann:2012fr} This is again a complex issue, as the extended spectral range of absorption may be overbalanced by recombination losses, or alternatively, by wasteful triplet formation.\cite{Deibel:2010do,Schlenker:2012ct}

\begin{figure*}[ht]
    \centering
        \includegraphics[width=\textwidth]{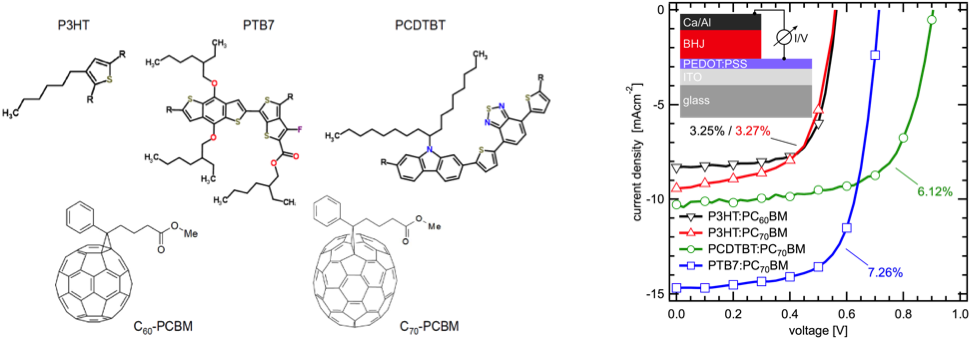}
   \caption{\textbf{Left.} Structures of the monomeric units of the polymers P3HT, PTB7, PCDTBT and the fullerene derivatives C$_{60}$-PCBM and C$_{70}$-PCBM. \textbf{Right.} Current-voltage (IV) characteristics of bulk heterojunction solar cells prepared from the polymer:fullerene composites shown on the left. Numbers state the achieved power conversion efficiency (PCE). The insert schematically illustrates the device structure. Details are provided in the Supporting Information.}
    \label{fig1}
\end{figure*}

Using experimental, EPR, and theoretical, DFT, methods we carried out a comparative study of electronic properties of cation and anion radicals in two polymers which are representative for major series of high-performance low band gap polymers: Poly[[9-(1-octylnonyl)-9H-carbazole-2,7-diyl]-2,5-thiophenediyl-2,1,3-benzothiadiazole-4,7-diyl-2,5-thiophenediyl] alias PCDTBT\,~\cite{Park:2009kb,Blouin:2008fu,Blouin:2007iy} and Poly[[4,8-bis[(2-ethylhexyl)oxy]benzo[1,2-b:4,5-b']dithiophene-2,6-diyl][3-fluoro-2-[(2-ethylhexyl)carbonyl]thieno[3,4-b]thiophenediyl]] alias PTB7\,~\cite{Liang:2010jb,Liang:2009ig}, and, as reference, the conventional polymer P3HT, in blends with either C$_{60}$-PCBM or C$_{70}$-PCBM (Figure~\ref{fig1}). To confirm the efficiency of the materials under study we prepared BHJ solar cells from the same polymer:fullerene composites as the EPR samples. The photovoltaic performance of the P3HT blended with C$_{60}$-PCBM or C$_{70}$-PCBM in comparison with the low band gap copolymers based solar cells is shown on Figure~\ref{fig1}. Here we report on the photo-induced charge separated state in polymer-fullerene composites. To our knowledge, this is the first EPR characterization of the charge-separated polaron states (radical cation and anion) in highly performing low band gap type polymer-fullerene BHJ. The use of the high resolution of high-frequency (HF) EPR at 130 GHz allows the unambiguous identification of the spectra of the polarons and the determination of its electronic $g$-tensor. The $g$-tensor is a sensitive probe of the electronic wave function and is influenced by the distribution of the unpaired electron on the polaron. To address the delocalization of the positive polaron on the polymer in more detail we also performed pulsed ENDOR experiments to determine the hyperfine (hf) interactions of magnetic nuclei with the unpaired electron. The key magnetic parameters, $g$-tensor and the hf-interaction are compared to the results of Density Functional Theory (DFT) calculations and correlated with the electronic properties of the polymer.

\section{Experimental}
\subsection{Sample Preparation}
The organic polymers studied in this work were highly regioregular P3HT (also known as P200), PCDTBT, and PTB7 (Figure~\ref{fig1}). P3HT was obtained from Rieke Metals (Lincoln, NE); PCDTBT and PTB7 were synthesized as described previously.\cite{Blouin:2008fu,Blouin:2007iy,Liang:2010jb,Liang:2009ig} The soluble fullerene derivatives C$_{60}$-PCBM and C$_{70}$-PCBM were obtained from Sigma-Aldrich (St. Louis, MO) and Solenne BV (Groningen, Netherlands). Polymer-fullerene mixtures (1:2 or 1:4 weight ratios) were prepared under anaerobic conditions in a N$_2$ dry box, using de-oxygenated toluene or chlorobenzene as solvents (Sigma-Aldrich). The concentrations of the solutions were in 5--15~mg/mL range. The solutions were filled into EPR quartz tubes, sealed under N$_2$-atmosphere, and then frozen quickly in liquid nitrogen. Films were prepared by slowly pumping the samples to remove the solvent gradually. Upon mixing of polymer and fullerene all further steps were performed under dimmed light. Both films and frozen solution were studied at cryogenic temperatures to minimize light-induced degradation (photodamage) of the sample.

Using the same polymer-fullerene composites as active layers, we fabricated three types of solar cells, which demonstrated power conversion efficiencies between 3.3 and 7.2\,\%, being representative for the respective material combinations. Figure~\ref{fig1} shows the current-voltage (IV) characteristics of these BHJ solar cells. The photovoltaic performance of the P3HT blended with either C$_{60}$-PCBM or C$_{70}$-PCBM is lowest as compared to the low band gap copolymers based BHJ. In P3HT:C$_{60}$-PCBM, the solar cells yield an open circuit voltage (V$_\text{OC}$) of 560~mV and short-circuit current of 8.31~mA/cm$^2$. Due to better absorption of C$_{70}$-PCBM as compared to C$_{60}$-PCBM, a slight increase in the short-circuit current J$_\text{SC}$ in the P3HT:C$_{70}$-PCBM device (9.43~mA/cm$^2$) along with a fill factor decrease was measured, thus yielding a similar power conversion efficiency (PCE) of $\sim$3.25\,\% for both devices. Using the copolymer PCDTBT together with C$_{70}$-PCBM in solar cells yields a much higher V$_\text{OC}$ of 905~mV and somewhat higher photocurrent, resulting in a PCE of 6.12\,\%. Finally, using PTB7:C$_{70}$-PCBM yields a  V$_\text{OC}$ of 718~mV, \textit{i.e.} in between those above two polymers, but much higher J$_\text{SC}$ of almost 15~mA/cm$^2$ and hence the PCE of 7.26\,\%, due to its superior optical absorption properties.\cite{Baumann:2012fr} See Supporting Information for a full data set of V$_\text{OC}$, J$_\text{SC}$, fill factor and PCE.

\subsection{EPR Spectroscopy}
Continuous wave (cw) X-band (9 GHz) EPR experiments were carried out with a Bruker ELEXSYS E580 EPR spectrometer (Bruker Biospin, Rheinstetten, Germany), equipped with a Flexline dielectric ring resonator (Bruker ER 4118X-MD5-W1) and a helium gas-flow cryostat (CF935, Oxford Instruments, UK). The temperature controller was an ITC (Oxford Instruments, UK). Pulsed EPR and pulsed ENDOR experiments were performed on the same spectrometer using a Bruker EN 4118X-MD4-W1 resonator and a BT01000-AlphaSA 1~kW RF amplifier (TOMCO Technologies, Stepney, Australia). ENDOR spectra were recorded using (i) the Davies ENDOR sequence\,~\cite{Davies:1974gu} ($\pi - t - \pi/2 - \tau - \pi - \tau - \text{echo}$) with an inversion $\pi$ pulse of 148~ns, $t = 10$~\textmu s, and radiofrequency (RF) $\pi$-pulse of 6~\textmu s, or (ii) the Mims ENDOR sequence\,~\cite{Mims:1965ck} ($\pi/2 - \tau - \pi/2 - t - \pi/2 - \tau - \text{echo}$) with $\pi/2$ pulse of 24~ns, $t = 10$~\textmu s, and RF $\pi$-pulse of $t = 6$~\textmu s.

Light excitation was done directly in the resonator with 532~nm Laser light (Nd:YAG Laser with OPO, model Vibrant from Opotek, operating at 10~Hz) or a 300~W Xenon lamp (LX~300F from Atlas Specialty Lighting with PS300-13 300~W power supply from Perkin Elmer). When using the lamp, a water filter (20~cm pathlength) was used to avoid unwanted heating of the sample. In addition, a KG3 short pass filter (Schott) removed residual IR irradiation. In both setups (Laser and lamp), a GG400 long pass filter (Schott) was used to remove UV light and hence to minimize photodamage of the sample. A lens (200~mm focal width) focused the light on the window in the resonator. Typical incident light intensities at the sample were around 2~W for the lamp and 40~mW for the Laser. 

High frequency (HF) EPR measurements were performed on a home-built D-band (130~GHz) spectrometer equipped with a single mode TE011 cylindrical cavity.\cite{Poluektov:2002be} EPR spectra of the samples were recorded in pulse mode in order to remove the microwave phase distortion due to fast-passage effects at low temperatures. The absorption line shape of the EPR spectra was recorded by monitoring the electron spin echo (ESE) intensity from a two microwave pulse sequence as a function of magnetic field. The duration of a $\pi/2$ microwave pulse was 40--60~ns, and typical separation times between microwave pulses were 150--300~ns. Light excitation was done directly in the cavity of the spectrometer with 532~nm Laser light through an optical fiber (Nd:YAG Laser, INDI, Newport, operating at 20~Hz, and OPO, basiScan, GWU). Typical incident light intensities at the sample were 20~mW. 

Data processing was done using Xepr (Bruker BioSpin, Rheinstetten) and MatlabTM 7.11.1 (MathWorks, Natick) environment. The magnetic parameters were obtained from theoretical simulation of the EPR and ENDOR spectra. Simulations were performed using the EasySpin software package (version 4.0.0) and the Kazan Viewer Software package.\cite{Stoll:2006ks,Silakov:2011td} Several EPR simulations were repeated with the program SimFonia Version 1.25 (Bruker BioSpin, Rheinstetten), using second order perturbation theory, and delivered virtually identical parameters. The relative accuracy in determination of the electronic $g$-tensor for the D-band EPR spectra is estimated to be $\pm 0.0001$. These values of the electronic $g$-tensor were taken as constraints for the X-band simulations.

\subsection{Density Functional Theory (DFT) Calculations}
Oligomers of various lengths were constructed for the three polymers P3HT, PCDTBT, and PTB7. The geometry optimizations were carried out using density functional theory (DFT) with the B3LYP functional\,~\cite{Becke:1993is,Stephens:1994jt,Lee:1988hx,Vosko:1980fd} using the 6-31G basis set, as implemented in PQSMol.\cite{Baker:2009gk} To test the effect of basis set and functional choice, geometry optimizations using the 6-31G$^*$ basis set or the BP86 functional were performed for several P3HT oligomers. Since all calculations led to nearly identical structures, the gas phase geometries as obtained with the B3LYP functional and the 6-31G basis set were used throughout in this study. Several DFT calculations which utilized the COSMO dielectric continuum model resulted in identical geometric structures and did not show a pronounced effect on the magnetic resonance parameters; thus, all calculations reported here were done in vacuo. This is also justified by the experimental approach, since nonpolar solvents were used in this study. After confirming by the absence of imaginary frequencies that the stationary points obtained in the geometry optimizations were minima, the spectroscopic parameters were obtained via single point DFT calculations, performed with the program package ORCA\,~\cite{Neese:2010vm} with the B3LYP functional in combination with the EPRII basis set\,~\cite{Rega:1996hw,Barone:1995kt}; sulfur was treated with the def2-TZVPP basis set of Ahlrich's and co-workers.\cite{Weigend:2005dh,Schafer:1992fv,Schafer:1994de} To test for influence of basis set on the calculated EPR parameters, several additional single point calculations were performed using the def2-TZVPP basis set for all atoms and showed only minor differences in the magnetic parameters. Hence, all calculated EPR parameters reported here were obtained with the EPRII basis set for all atoms except sulfur. The Supporting Information contains a comparison of calculations using different basis set and sidechains for the three polymers. The principal $g$-values were calculated employing the coupled-perturbed Kohn-Sham equations,\cite{Neese:2001kh} in conjunction with a parameterized one electron spin-orbit operator. The anisotropic magnetic dipole and the isotropic Fermi contact contributions to the hf-coupling were calculated for all $^1$H, $^{14}$N and $^{19}$F atoms. Second-order spin-orbit contributions were found to be negligible for these light atoms.

\section{Results and Discussion}
\subsection{EPR Spectroscopy}
\begin{figure*}
    \centering
    \includegraphics[width=.8\textwidth]{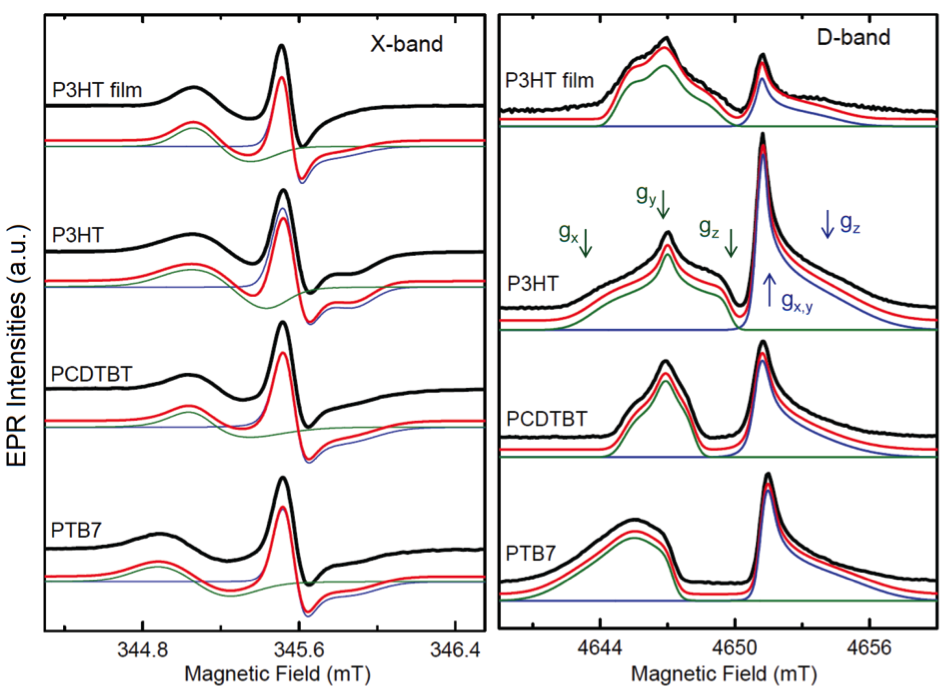}
   \caption{Light-induced EPR spectra of polymer-fullerene blends (composites) in frozen toluene solution, containing P3HT, PCDTBT, or PTB7 and C$_{60}$-PCBM. In addition, light-induced EPR spectra of a P3HT:C$_{60}$-PCBM film are shown on top. All measurements were done at 50~K. \textbf{Left.} CW X-band EPR spectra, resulting in derivative line shape. \textbf{Right.} Echo-detected field-swept D-band spectra, resulting in absorptive line shape. Black: Experimental spectra. Green: Simulation of the positive polaron on the polymer. Positions of the $g$-tensor principal components are shown by green arrows. Blue: Simulation of the negative polaron on the fullerene. Positions of the $g$-tensor principal components are shown by blue arrows. Red: Sum of the two simulations. Note, that the positions of the spectra were corrected for small differences in MW frequency to allow direct comparison. The principal $g$-values for the simulation are provided in Table~\ref{tab1}. EPR spectra of the polymer cation in the film were simulated using effective $g$-values and preferential ordering around the X-axis (EasySpin built-in ordering function; 0.44).}
    \label{fig2}
\end{figure*}

\begin{figure*}
    \centering
    \includegraphics[width=.8\textwidth]{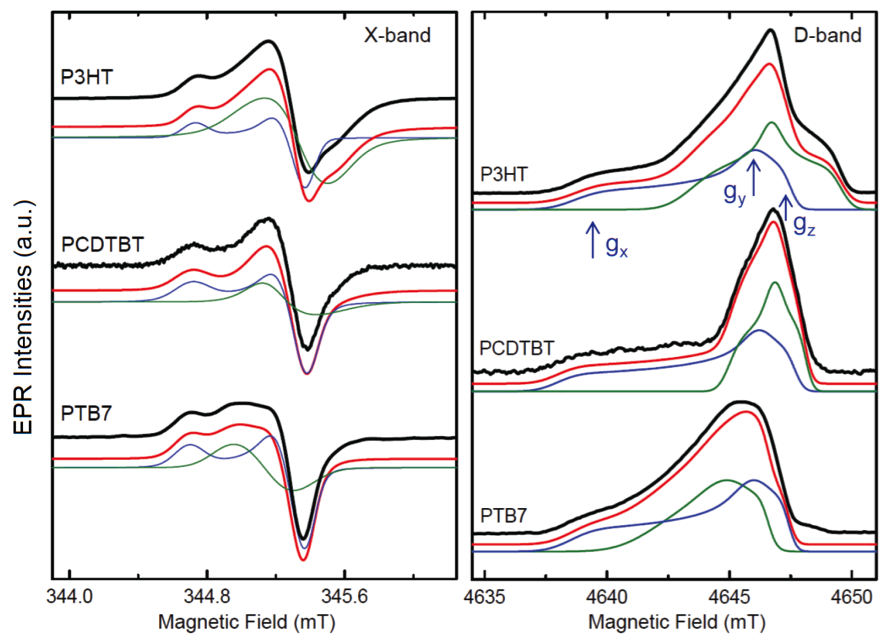}
   \caption{Light-induced EPR spectra of polymer-fullerene blends (composites) in frozen solution of toluene, containing P3HT, PCDTBT, or PTB7 and C$_{70}$-PCBM. All measurements were done at 50~K. \textbf{Left.} CW X-band spectra, resulting in derivative line shape. \textbf{Right.} Echo-detected field-swept D-band spectra, resulting in absorptive line shape. Black: Experimental spectrum. Green: Simulation of the positive polaron on the polymer. Blue: Simulation of the negative polaron on the fullerene. Positions of the $g$-tensor principal components are shown by blue arrows. Red: Sum of the two simulations. Note, that the positions of the spectra were corrected for small differences in MW frequency to allow direct comparison. }
    \label{fig3}
\end{figure*}

Before illumination, only very weak EPR signals were observed in all three different composites/blends, which we attribute to traces of paramagnetic impurities and a minor number of molecules already in a charged state (data not shown). Upon illumination, intense EPR signals around $g \approx g_e$ were detected (Figures~\ref{fig2}~\&~\ref{fig3}). Special precautions were taken to minimize the presence of molecular oxygen in the samples during the experiments. While the presence of oxygen in the dark did not generate any EPR signals, the presence of even small amounts of oxygen during illumination leads to the appearance of additional EPR signals, which are attributed to defects on the polymer and radical complexes with oxygen.  After extended illumination, degradation and complete shutdown of charge separation efficiency were observed, resulting in stable EPR signals. These findings are in line with studies on photodamage in OPV materials.\cite{Aguirre:2011ic,Cook:2012hw,Jorgensen:2011fk,Sperlich:2011ug} P3HT:fullerene blends demonstrate somewhat higher stability compared to PCDTBT and PTB7 composites. Therefore, for these blends, it was essential to perform the measurements below 100~K, which allows carrying out experiments for several hours. 

First, we will discuss the composites which contained C$_{60}$-PCBM as electron acceptor (Figure~\ref{fig2}). Two different signals were detected both at X-band and D-band. Following previous studies, these two signals can be unequivocally assigned: the low field signal stems from the positive polaron on the polymer, while the high field signal is due to the negative polaron on the fullerene cage, in this case C$_{60}$-PCBM.\cite{Ceuster:2001hc,Poluektov:2010ie,Aguirre:2008bw} At 50~K, where these measurements were performed, a significant fraction of the two signals is persistent on the timescale of hours after switching the light off. In frozen solution, the major part of the signals is stable, while in films up to 70\,\% of the signals are reversible. This difference can be explained by the smaller average distance between polymer and fullerene molecules in films, which facilitates charge recombination. Annealing the samples under anaerobic conditions in the dark at room temperature resulted in the disappearance of the light-induced EPR signal. At X-band frequency (Figure~\ref{fig2}, left), the signals of the positive and the negative polaron partially overlap and the three principal values of the $g$-tensor are not resolved. This can be clearly seen in the simulation of the individual polarons. To separate the overlapping signals and to determine unambiguously the electronic $g$-tensor, we used HF EPR spectroscopy. The light-induced EPR spectra at 130 GHz (Figure~\ref{fig2}, right) show both a clear separation of the positive polaron on the polymer (low field side) and the negative polaron on the fullerene (high field side). Furthermore, for all three polymers, the three principal components of the electronic $g$-tensor are resolved. The principal values of the $g$-tensors for the positive polaron on P3HT, PCDTBT, and PTB7 are summarized in Table~\ref{tab1}. The $g$-tensor of P3HT agrees with a previous study within the error of our measurements, $\pm$0.0001.\cite{Poluektov:2010ie} To the best of our knowledge, this is the first determination of the electronic $g$-tensor of PCDTBT and PTB7. For both, in contrast to P3HT which has $g_z = 2.0011$, we observed no $g$-tensor component significantly lower than the free electron $g$-value ($g_e \approx 2.0023$).\cite{Mohr:2005il} In particular, PTB7 is clearly shifted to higher $g$-values, lower magnetic fields. The simulation of the negative polaron on C$_{60}$-PCBM showed an almost axially symmetric $g$-tensor with a very broad parallel component at high field. The $g$-values of the negative polaron on C$_{60}$-PCBM have been determined to be $g_x=2.0006$, $g_y=2.0005$, $g_z=1.9985$. The relative values are in good agreement with previous HF EPR studies, while the deviation of absolute values in the different studies can be attributed to calibration with $g$-standards.\cite{Ceuster:2001hc,Poluektov:2010ie,Aguirre:2008bw}

\begin{table}[ht]
\caption{The principal values of the $g$-tensors of positive, P$^+$, polarons on the polymers and negative, P$^-$, on the fullerenes, in polymer-fullerenes blends as determined experimentally in frozen toluene solution. The values given in parentheses are the effective $g$-tensor values for the film.}
\begin{center}
\begin{tabular}{l|cccccc}
 & P$^+$ in  & P$^+$ in  & P$^+$ in & P$^-$ in & P$^-$ in  \\ 
$g$$^{ ~a)}$ & P3HT  & PCDTBT & PTB7 &  C$_{60}$-PCBM &C$_{70}$-PCBM \\ 
\hline
$g_{x}$   &  2.0038 (2.0032)  &  2.0032  &  2.0045  &  2.0006  &  2.0060   \\
$g_{y}$   &  2.0023 (2.0023)  &  2.0024  &  2.0031  &  2.0005  &  2.0028   \\
$g_{z}$   &  2.0011 (2.0012)  &  2.0018  &  2.0023  &  1.9985  &  2.0021   \\
\hline
\end{tabular}
$^{a)}$ Relative error in the $g$-tensor measurements is $\pm$0.0001. The large distribution of the parallel component of the fullerene $g$-tensor induces an error of $\pm$0.0003.
\end{center}
\label{tab1}
\end{table}%

In general, a much larger line width at D-band than at X-band was detected. This holds both for the polymers and the fullerenes. Since the samples were prepared from identical solutions, we attribute this line broadening at D-band to $g$-strain: the narrow distribution of $g$-values due to subtle differences in the direct surrounding. The effect of $g$-strain on the line width is proportional to the microwave (MW) frequency, and thus $\sim$14 times more pronounced at D-band than at X-band. The principal values of the electronic $g$-tensor as determined by simulation of the D-band spectra were taken as constraints for the X-band simulations.

In the other two blends with PCDTBT or PTB7, the spectra of the negative polaron on C$_{60}$-PCBM are identical to the one in the blend with P3HT, indicating negligible interactions between polymer and fullerene in all cases. Note that EPR signals of polymers and fullerenes recorded in films (as shown for P3HT in Figure~\ref{fig2}) are somewhat more narrow as compared to the signals recorded in the frozen solutions, while the electron spin relaxation times of these signals (as measured by pulsed techniques) are similar. The narrowing is seen both for X-band and D-band. We attribute this effect to partial ordering of molecules on the walls of EPR tubes, which act as a substrate for the film. Other potential explanations: (i) higher delocalization of the unpaired electron spin in the film, or (ii) strongly interacting (close-by) electron spins were excluded based on the following reasoning. If the narrowing of EPR lines are due to higher delocalization of the unpaired electron spin in films, then the largest hyperfine couplings should decrease and similar narrowing should be seen in the $^1$H ENDOR spectra. However, the maximal observed $^1$H hyperfine coupling in the polymer is the same for frozen solutions of different concentrations and films (see ENDOR section below). If the reason for the narrowing of EPR lines is strongly interacting (close-by) electron spins, it would have a significant effect on the electron spin relaxation time, which was not detected by pulsed EPR relaxation measurements (data not shown) and can thus be excluded as well.

Due to the ordering effect in the P3HT:C$_{60}$-PCBM film the anisotropy of the $g$-tensor is reduced,\cite{Berliner:1976ug,Gaffney:1974dk} resulting in effective $g$-tensor values as given in Table~\ref{tab1}. Substantial ordering effects in polymers have been reported previously in the films of different OPV material blends.\cite{Aguirre:2008bw,Konkin:2009fy,Szarko:2010em} We did not observe orientation effects for the more spherical fullerenes. The $g$-values of the fullerene in the films are identical to those measured in solution. Slight changes in the line width of the C$_{60}$-PCBM anion in X-band EPR spectra can be explained in the following way: Narrowing of the perpendicular (low field) line in the film (0.10 vs. 0.13~mT) is due to absence of magnetic nuclei of the solvent as discussed in more detail in the ENDOR section below. On the other hand, the parallel (high field) component of the EPR spectra in the films is much broader than in any solvents (see Figure~\ref{fig2}). We attribute this to the $g$-strain effect (distribution of surrounding), which is already significant in the solvents and should be even larger in the disordered solid state. In light of this, to avoid the orientation effects, all measurements of $g$-tensors of anions and cations were done for very concentrated solutions, not oriented films. 

Figure~\ref{fig3} presents the X-band and D-band EPR spectra of the blends of the same three polymers with C$_{70}$-PCBM. The larger $g$-values of the negative polaron on C$_{70}$-PCBM as compared to C$_{60}$-PCBM result in strong overlap of the individual EPR spectra of positive and negative polarons. Similarly, this overlap is also present in the D-band spectra. However, all spectra could be simulated with the same magnetic parameters for the positive polaron independent of the fullerene used in the blend. We could also simulate the spectral contribution of negative polarons on C$_{70}$-PCBM with very similar parameters for the different blends. The principal $g$-values are in agreement with previously published values.\cite{Poluektov:2010ie} In C$_{70}$-PCBM, the largest $g$-tensor component (low field) exhibits a large $g$-strain effect, indicating the sensitivity on its direct surrounding. Slight variations of this $g$-value, in the order of $\pm 0.0003$, were observed in different samples. The use of chlorobenzene instead of toluene as solvent also had a comparable effect on this $g$-value. The Supporting Information shows a direct comparison of PCDTBT:C$_{70}$-PCBM blends in chlorobenzene and toluene.

It is important to emphasize that EPR signals shown on Figures~\ref{fig2} \& \ref{fig3} correspond to fully separated charge states. This is supported by the absence of any observable electron spin--spin interactions, \textit{i.e.} line broadening or appearance of low and high field extra signals due to strong electron--electron interactions. The X-band EPR line widths of the polymer cation and C$_{60}$-PCBM or C$_{70}$-PCBM anion do not depend upon the concentration of the polymer:fullerene blend in the solution. The line width of the individual components in X-band EPR spectra are 0.1--0.4~mT. Computer modeling demonstrates that 0.1--0.2~mT Gaussian broadening of these lines can be reliably detected in the EPR spectra. Note, that 0.1--0.2~mT broadening corresponds to the dipole--dipole interaction between two electrons separated by a distance in the order of 25--30~\AA ~(depending upon the inter-spin orientation in respect to the external magnetic field). Thus, taking into account our previous discussion on the line width in the films, we can claim that under our experimental conditions the recorded radicals are separated from each other at least for 25~\AA.

\subsection{ENDOR Spectroscopy}

\begin{figure}
    \centering
    \includegraphics[width=.45\textwidth]{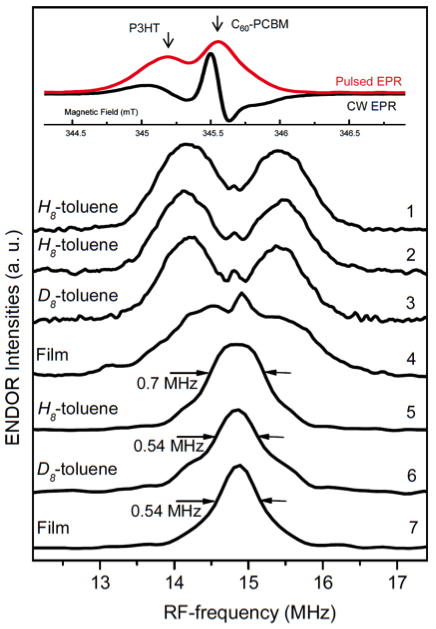}
   \caption{Light-induced X-band ENDOR spectra of P3HT:C$_{60}$-PCBM polymer-fullerene blends. Measurements were done at 50~K. 1, 2, 3, 4 -- were measured at the maximum of P3HT signal; 5, 6, 7 -- were measured at the maximum of C$_{60}$-PCBM signal. 1, 5, 6, 7 -- Mims-type ENDOR; 2, 3, 4 -- Davies-type ENDOR. \textbf{Insert.} Corresponding cw and echo-detected field-swept X-band spectra of P3HT:C$_{60}$-PCBM polymer:fullerene blends in H$_8$-toluene. The spectral positions where the ENDOR experiments have been performed are marked by arrows.}
    \label{fig4}
\end{figure}

While the D-band experiments allow a precise determination of the $g$-tensor, the hyperfine interaction of the unpaired electron with magnetic nuclei ($I \neq 0$) is neither resolved at D-band nor X-band frequency. This hf-interaction is a magnetic interaction between the spin of the unpaired electron and magnetic nuclei. The magnetic nuclei within these composites are $^1$H for all three polymers and two fullerenes, as well as $^{14}$N (for PCDTBT), and $^{19}$F (for PTB7). It is the sum of an isotropic contribution (the Fermi contact interaction) and an anisotropic  (dipolar) contribution. A larger amount of spin density at the position of the respective nucleus results in a larger isotropic hf-interaction, while a larger amount of spin density close to the respective nucleus results in a larger dipolar, anisotropic hf-interaction. A larger delocalization over more monomeric units thus decreases the largest detected hf-coupling, but increases the number of nuclei coupled to the unpaired electron spin. For example, the hf-interaction of the thiophene ring proton will, to first order, decrease as $1/n$, where $n$ is the number of monomeric units over which the polaron is distributed. The hf-couplings determined with the ENDOR experiments can then be compared with the results of DFT calculations on oligomers of different length, where the length of the oligomer presents an upper bound for the delocalization length. The experimental results provide in this case a clear reference point for the theoretical calculations. Hence, pulsed ENDOR experiments\,~\cite{Davies:1974gu,Mims:1965ck,Gemperle:1991dm} at X-band were performed which allow the direct measurement of these hf-interactions. Davies- and Mims-type\,~\cite{Davies:1974gu,Mims:1965ck} ENDOR were detected on the maximum of the pulsed EPR signals of the cation on the respective polymer and anion of C$_{60}$-PCBM, as indicated by arrows in Figure~\ref{fig4} (upper part).

In the case of P3HT, the ENDOR signal, as shown in Figure~\ref{fig4}, is centered around the proton Larmor frequency of 14.8~MHz and exhibits an overall spread of about 2.5~MHz. The spectra recorded with two different pulse sequences, Davies- and Mims-type ENDOR, are very similar (compare Figure~\ref{fig4}.1 and ~\ref{fig4}.2). The Mims-type ENDOR spectra demonstrate better S/N ratio as Mims technique is more sensitive to the small couplings compared to the Davies-type ENDOR. We used both fully protonated toluene (H$_8$-toluene) and fully deuterated toluene (D$_8$-toluene) as a solvent. Both ENDOR spectra are super-imposable, indicating negligible interaction with the surrounding solvent matrix. In film, the maximal hyperfine interaction matches the one measured in frozen toluene solution, indicating an identical electronic structure and degree of delocalization of the polaron in the polymer. For both PCDTBT and PTB7 larger $^1$H hf-couplings were observed, up to 4~MHz and 6~MHz respectively (see Figure~\ref{fig7}). No ENDOR signals were detected for $^{14}$N in the blend with PCDTBT. ENDOR signals of $^{19}$F in PTB7:fullerene blends were not resolved due to strong overlap with more intense $^1$H signal. A detailed analysis and discussion of the $^1$H hf-couplings of the positive polaron can be found below based on the DFT results.

\newpage
The Mims-type ENDOR spectra of the C$_{60}$-PCBM anion were recorded in the film and in the concentrated solutions of H$_8$-toluene and D$_8$-toluene, Figure~\ref{fig4}.5--\ref{fig4}.7. These spectra consist of a narrow line in the center of the spectra and broad wings. In the P3HT:C$_{60}$-PCBM blend, the individual EPR signals of the positive and negative polaron partially overlap, which manifests itself as the broad wings in the C$_{60}$-PCBM ENDOR spectra. The narrow line has the same width in the film and D$_8$-toluene solution, 0.54~MHz, while in the toluene H$_8$-toluene solution it is slightly larger, 0.70~MHz. This demonstrates that the main source of hf-coupling in the film is interaction with distant protons of fullerene and polymer side chains. The narrowing of the ENDOR signal of the C$_{60}$-PCBM anion is also consistent with the decrease of EPR line width (0.13~mT in H$_8$-toluene vs. 0.10~mT in D$_8$-toluene and films).

\subsection{DFT Calculations and Comparison with Experimental Data}

DFT calculations were performed to determine the geometric and electronic structure of the respective polymer in its singly oxidized state, \textit{i.e.} as a radical cation. This is the oxidation state of the polymer after donating an electron to an electron acceptor, like a fullerene molecule. Of particular interest is the extent of delocalization of the positive polaron and the resulting magnetic resonance parameters. The comparison of calculated magnetic resonance parameters with those obtained experimentally provides an excellent way to verify the extent of charge delocalization in the polymer.

\begin{figure}
    \centering
    \includegraphics[width=.45\textwidth]{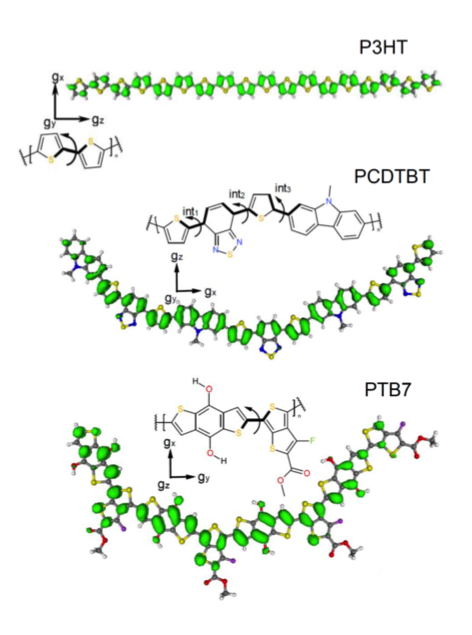}
   \caption{Spin density isosurface plots of the lowest energy conformation of the largest oligomer calculated for each of the polymers studied. The arrows indicate the bonds which allow internal rotations. Note, that for all three polymers other conformers with similar energies exist. Spin density isosurface plots of selected geometries, relative energies and magnetic parameters of some conformers are provided in the Supporting Information. All isosurfaces are shown at a contour level of $0.0005~e/a_0^3$. The orientation of the principal axes of the electronic $g$-tensor is also given. \textbf{Top.} P3HT 18-mer. Dihedral angles are all 180$^\circ$ (``trans'' conformer). \textbf{Middle.} PCDTBT trimer. The three internal dihedral angles (int$_1$-int$_3$) are 0$^\circ$, 0$^\circ$, and 0$^\circ$ in each monomeric unit, and the external angles are all 0$^\circ$. \textbf{Bottom.} PTB7 tetramer. The internal dihedral angle is 180$^\circ$ in each monomeric unit, and the external angles are all 0$^\circ$.}
    \label{fig5}
\end{figure}

Oligomers of various lengths were constructed for the three polymers P3HT, PCDTBT, and PTB7. Due to computational constraints, the maximum number of monomeric units of P3HT oligomers was 18, for PCDTBT three, and for PTB7 four. In all calculations, the side chains were truncated: in P3HT to H; in PCDTBT to methyl groups; in PTB7 the side chains were truncated to H on ether groups and to methyl on carbonyl groups (compare Figure~\ref{fig1} with Figure~\ref{fig5}). Test calculations using more complete side chains for certain short oligomers (see Supporting Information) indicate that this truncation did not fundamentally affect the calculated results for the geometries, $g$-values, or spin densities, unless very small oligomers were used. Additional calculations were done on a stacked dimer of trimeric P3HT oligomers, since stacked dimers may form under certain experimental conditions, like in very concentrated solutions or solid films. Here, the sidechains were truncated to ethyl groups to account for possible intermolecular interactions.

Figure~\ref{fig5} shows the lowest energy conformation of the largest oligomer calculated for P3HT, PCDTBT, and PTB7. For P3HT, the calculations were performed for oligomers ranging from the monomer to the 9-mer, 12-mer, 13-mer, 15-mer and 18-mer oligomers. In the case of P3HT, geometry optimizations always yielded planar structures (referring to the thiophene unit). Note, that the rotational barrier is higher for the cation radical than for the neutral P3HT oligomer (20~kcal/mol vs. 4~kcal/mol). For all but the shortest P3HT oligomers, the vast majority of the unpaired spin density is distributed over the conjugated thiophene rings, with virtually no spin density on the alkyl side chains (see Supporting Information). The absence of spin density on sulfur atoms is in agreement with the experimental results, since a significant amount of spin density on the sulfur atoms would have resulted in a larger $g$-tensor anisotropy due to the larger spin-orbit coupling constant of sulfur. The middle $g$-value ($g_y$) closely resembles $g_e$ for all calculated oligomer lengths (see Supporting Information), and is oriented perpendicular to the plane of the \mbox{$\pi$-system} (Figure~\ref{fig5}). The $g_z$-value is lower than $g_e$ and is oriented along the backbone of the oligomer chain. Both $g_y$ and $g_z$ values are in reasonable agreement with experimental values and do not depend systematically on the length of the oligomer (number of monomeric units) or the conformation. The $g_x$ value (which lies in the plane of the \mbox{$\pi$-system} and perpendicular to the axis of the monomeric units) is more sensitive to the length of the oligomer and to the type of the sidechain (see Supporting Information). Based on the DFT calculations a localization of the positive polaron on a monomer can be excluded. However, a definitive length for delocalization cannot be determined from the $g$-values alone due to the weak dependence of the calculated $g$-tensor on the oligomer length which is comparable with the accuracy of these DFT calculations.

\begin{figure}
    \centering
    \includegraphics[width=.4\textwidth]{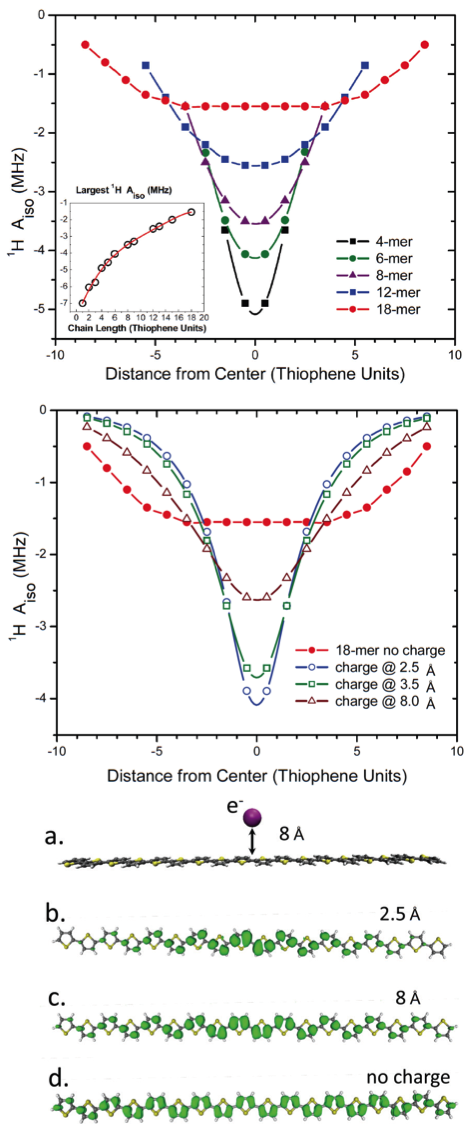}
   \caption{\textbf{Top.} Calculated isotropic $^1$H hyperfine coupling constants for the hydrogen atoms of the positively charged thiophene units as a function of the P3HT oligomer length. Black: 4-mer; Green: 6-mer; Magenta: 8-mer; Blue: 12-mer; Red: 18-mer. Insert. Dependence of the maximum absolute value of the isotropic $^1$H hf-coupling as a function of P3HT oligomer length. 
\textbf{Middle.} Calculated isotropic $^1$H hyperfine coupling constants for the hydrogen atoms of the positively charged thiophene units for an 18-mer with a negative point charge placed in the center of the oligomer at a distance of 2.5~\AA ~(blue), 3.5~\AA ~(green), 8~\AA ~(magenta), and without countercharge (red). 
\textbf{Bottom.} Geometry of a positively charged P3HT 18-mer with one negative charge placed away from the center of the oligomer \textbf{(a)} and Spin density isosurface plot of a P3HT 18-mer with negative charge at: \textbf{(b)} -- 2.5~\AA,  \textbf{(c)} -- 8~\AA, \textbf{(d)} -- no negative charges in the vicinity.
}
    \label{fig6}
\end{figure}

The isotropic hf-interaction of the unpaired electron with magnetic nuclei, however, is proportional to the unpaired spin density at the respective nucleus and should provide a more reliable measure of the degree of delocalization. For these calculations, the alkyl sidechains on P3HT were replaced by hydrogen atoms to allow longer oligomers to be included. Figure~\ref{fig6} shows the calculated $^1$H isotropic hf-coupling constants averaged over each thiophene unit as a function of distance from the center of the thiophene chain. As the overall length of the oligomer increases, the maximum magnitude of the coupling decreases for hydrogen atoms located on the central part of the chain. Additionally, the values almost become constant over a large part of the chain. The dipolar part of the hf-coupling tensor demonstrates a similar dependence as the isotropic part (see Supporting Information). Thus, both the isotropic and dipolar terms reveal clear sensitivity to the oligomer length. 

\begin{figure}
    \centering
    \includegraphics[width=.45\textwidth]{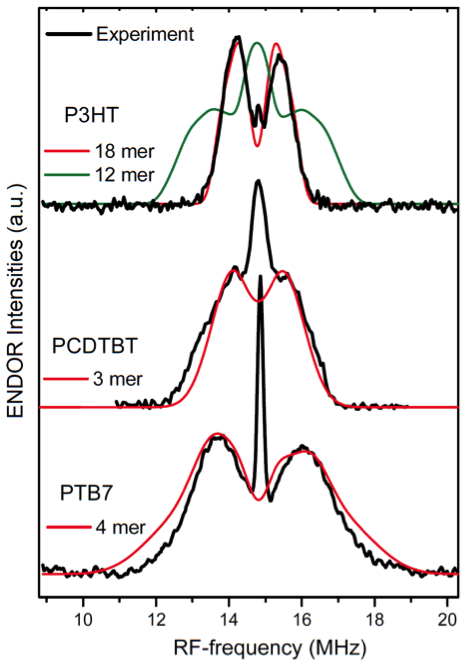}
   \caption{Light-induced pulsed ENDOR spectra of polymer:C$_{60}$-PCBM blends. Black -- experiment. Red -- best simulation based on DFT modeling done for: 18-mer of P3HT cation; 3-mer for PCDTBT cation; 4-mer for PTB7 cation (for details see Supporting Information). Green -- simulation based on DFT modeling of 12-mer P3HT cation, which does not agree with experimental data.}
    \label{fig7}
\end{figure}

In order to determine the delocalization length of the positive polaron in P3HT, ENDOR spectra were simulated using the $^1$H hf-coupling as obtained from DFT calculations on oligomers of different length. Several of these simulations are shown in Figure~\ref{fig7}.

In the cases of the monomer, tetramer, 8-mer, and 12-mer the line width of the theoretically simulated ENDOR spectra substantially exceeds the experimental one, while the ENDOR spectrum simulated for the 18-mer is in excellent agreement with experiment (Figure~\ref{fig7}). Note, that for these oligomers the sidechains were omitted from the calculations. In order to justify this truncation, we simulated two sets ENDOR spectra for the 4-mer and 8-mer, one of which with the complete hexyl sidechain and one with hydrogen replacing the sidechain (see Supporting Information). The similarity of the width of these spectra confirms that this simplified treatment is appropriate. Based on the analysis of the hf-tensor dependence on the length of oligomer (Figure~\ref{fig6} \& Supporting Information), we conclude that the delocalization length of positive polaron in P3HT is at least 15~units, which corresponds to ca. 60~\AA. We define the delocalization length as the distance at which the absolute value of the isotropic $^1$H hf-coupling drops to half of the maximum one (FWHM of the curves on Figure~\ref{fig6}). Prior theoretical treatment\,~\cite{Geskin:2003bl} significantly underestimates the delocalization length to be around 6 thiophene units. Geskin et al. used the conjugation length (C--C bond length changes) and Mulliken charge distributions as measures of the delocalization. Our DFT calculation of a 12 thiophene long oligomer, using the B3LYP functional and the EPRII basis set for light elements, delivers a similar C--C bond distance distribution as in ref.~\cite{Geskin:2003bl}. The more crucial difference between this work and prior work is the choice of functional. As noted by Nayyar et al.~\cite{Nayyar:2011hv} there is a correlation between the degree of localization and the percentage of Hartree-Fock component in the functional. Consequently, the BHandHLYP functional (50\,\% Hartree-Fock), used by Geskin et al.~\cite{Geskin:2003bl}, predicts significantly more localized spin density than the B3LYP functional (20\,\% Hartree-Fock). However, the simulated ENDOR spectrum utilizing calculated hf-coupling constants of shorter oligomers does not agree with the experimental one (Figure~\ref{fig7}). We thus performed DFT calculations on larger oligomers and only the 18-mer yielded hf-couplings in agreement with our experimental results. In fact, we believe that the prior agreement between the more localized polaron of the BHandHLYP functional calculations and experimental work is due to extra charged species causing the localization and that the true picture is a delocalized one.

The total spread of the ENDOR spectrum of the P3HT polaron is 2.5~MHz, which is significantly narrower than reported previously.~\cite{Aguirre:2008bw} In the latter ENDOR and HYSCORE study a maximum hf-coupling of 4.6~MHz was extracted from the experiment. This can be attributed to a stronger localization of the positive polaron over the polymer chain compared to our study. The most likely reason for this discrepancy is the iodine treatment used to generate the charged polarons on the polymer chain in ref.~\cite{Aguirre:2008bw}. The chemical modification of the polymer chain and/or possible presence of the I$^-$ (or I$_3^-$) counter ion close to the polymer chain may well be a cause for stronger localization of the polaron on the polymer. To verify this assumption we performed DFT calculation of positive charge distribution in P3HT 18-mer in the presence of a close by negative charge as a function of distance between the negative charge and the polymer chain. The calculations reveal that a negative counter charge in the vicinity of the polymer chain substantially narrows the distribution of the positive polaron along the polymer chain (see Figure~\ref{fig6}). The light-induced oxidation method employed in this study does not require the addition of an oxidant which functions as counter-ion, and thus more closely mimics the conditions present in functional BHJ cells.

Similar analyses of the experimental data in combination with DFT calculations have been done for the cases of PCDTBT and PTB7. For PCDTBT, the calculations were performed for monomer, dimer, and trimer oligomers. Since the majority of the unpaired spin density is distributed over the conjugated rings, the ethyl-hexyl side chains were replaced by methyl groups. These replacements did not noticeably affect structure or the $g$-tensor values (see Supporting Information). The geometry optimizations yielded planar structures (referring to the various rings). Several of different PCDTBT conformers show only small differences in energy. The rotational barriers (internal and external rotations as shown in Figure~\ref{fig5}) were found to be in the range of 3--6~kcal/mol for neutral PCDTBT. Thus, a mixture of conformers may be present in the sample studied by EPR spectroscopy. We therefore calculated the magnetic parameters for a variety of conformers (see Supporting Information). Even for the monomer there is good agreement between the calculated $g$-tensor values (2.0010--2.0019, 2.0023, 2.0029--2.0037) and the experimental values (2.0018, 2.0024, 2.0032). The middle $g$-value ($g_y$) closely resembles $g_e$ for all calculated oligomer lengths (see Supporting Information), and is oriented perpendicular to the plane of the \mbox{$\pi$-system} (Figure~\ref{fig5}). The $g_z$ value is slightly lower than $g_e$ and is oriented perpendicular the backbone of the oligomer chain. Both $g_y$ and $g_z$ values are in reasonable agreement with experimental values and do not depend systematically on the length of the oligomer (number of monomeric units) or the conformation. The $g_x$ value (which lies in the plane of the \mbox{$\pi$-system} and parallel to the axis of the monomeric units) is more sensitive to the length of the oligomer. Similar to the P3HT case, the increase of the oligomer size has a negligible effect on the $g$-tensor, and does not allow the determination of the extent of delocalization of the polaron. As for P3HT, the isotropic hf-interaction of the unpaired electron with magnetic nuclei was used to estimate the degree of delocalization. Since the monomeric unit is much larger than in P3HT, the model oligomer could only be increased in larger steps. The width of the experimental ENDOR spectra is about 4~MHz. Simulations using the magnetic parameters from the DFT calculations on monomer and dimer could not reproduce the width of the experimental ENDOR spectrum, while the simulation using the hf-coupling constants of the trimer is similar to the experiment (Figure~\ref{fig7}). Since we could not perform calculations on larger PCDTBT oligomers, we conclude that the trimer represents a reasonable minimal model size for polaron delocalization in PCDTBT. 

For PTB7, calculations were performed on the monomer, dimer, trimer, and tetramer. While the component of the $g$-tensor oriented perpendicular to the \mbox{$\pi$-plane} ($g_z$) is not influenced by the length of the oligomer, the two in-plane components ($g_x$, $g_y$) generally decrease as the chain length increases. The largest and most sensitive component ($g_x$) is oriented perpendicular to the intrachain axis. The monomer model can be excluded on the basis of disagreement of the calculated vs. experimentally determined $g$-tensor. For the larger models, the calculations find several conformers with similar energetics but slightly different $g_x$-values (Supporting Information). The rotational barriers (internal and external rotations as shown in Figure~\ref{fig5}) were found to be in a range similar to PCDTBT. Hence, we expect that also here a mixture of conformers will be present in the sample studied by EPR spectroscopy. The calculated $g$-values of the tetramer models (2.0024, 2.0028--2.0034, 2.0040--2.0057) are in good agreement with the experimental one (2.0023, 2.0031, 2.0045). The calculated $g_x$-values vary as much as 0.0017, which is also in agreement with the experimental results: at D-band the low-field edge is significantly broadened. This broadening can be explained by $g$-strain effect of the $g_x$-component and can be estimated from the simulation of the EPR spectra as 0.0015. As for P3HT and PCDTBT, the isotropic hf-interaction of the unpaired electron with magnetic nuclei was used to estimate the degree of delocalization. Since the monomeric unit is much larger than in P3HT, the model oligomer could only be increased in larger steps. The width of the experimental ENDOR spectra is about 6~MHz. Simulations using the magnetic parameters from the DFT calculations on monomer and dimer could not reproduce the width of the experimental ENDOR spectrum, while the simulation using the hf-coupling constants of the trimer or tetramer models are similar to the experimental data with the latter one being in best agreement with the experiment (Figure~\ref{fig7}). We thus conclude that the tetramer represents a reasonable model size for polaron delocalization in PTB7.  

To summarize, for all the polymer cation radicals the calculated $g$-tensors are very close to experimental ones, if more than minimal size model oligomers were used. Hence, in all cases the calculated dependence of the \mbox{$g$-values} on the length of oligomers and types of conformers is not decisive for determining the polaron delocalization. However, the length of polaron delocalization in P3HT, PCDTBT and PTB7 could be determined by analysis of the ENDOR data in combination with the DFT calculation of $^1$H hf-couplings. The width of the experimental ENDOR spectra increases in the order P3HT, PCDTBT, PTB7 from 2.5~MHz to 4~MHz and 6~MHz, respectively (Figure~\ref{fig7}). The increased width of the ENDOR spectra indicates the increase of hf-coupling constants on the central, strongly coupled protons. This increase is connected to the decrease of the extent of polaron delocalization. Based on the DFT calculations of the $^1$H hf-couplings, the ENDOR spectra were simulated for PCDTBT and PTB7 oligomers of different lengths. The best fits were obtained for the trimer of PCDTBT and the tetramer of PTB7 (Figure~\ref{fig7}). The width of the simulated ENDOR spectra for smaller oligomers exceeds the experimental ones. This analysis shows that for PCDTBT the polaron is approximately delocalized along a length of ca. 50~\AA, while for PTB7 the delocalization length is ca. 40~\AA. 

The extent of the polaron delocalization along the polymer is considered crucial for efficiency of charge separation and charge recombination processes in OPV materials. A strong delocalization of the positive charge along the polymer chains leads to the substantial decrease of the Coulomb binding energy in the CT-state thus easing the primary charge separation. This mechanism of CT-state decay was recently discussed.\cite{Bakulin:2012if} 

It is interesting to compare the length of polaron delocalization on the polymer chain with power conversion efficiency of the BHJ cells prepared from the respective polymer:fullerene composites (Figure~\ref{fig1}). While the length of delocalization is increasing in the series PTB7, PCDTBT, P3HT as ca. 40~\AA, 50~\AA, 60~\AA, the conversion efficiency is decreasing in the same sequence as 7.2\,\%, 6.1\,\%, and 3.2\,\%. Therefore the increase of the polaron delocalization length does not imply higher conversion efficiency. On the one hand, the longer the delocalization the more efficient is the charge separation. On the other hand, wide-ranging delocalization leads to the more efficient recombination of charges as the overlap between donor and acceptor wavefunctions will increase. Evidently the delocalization should be optimal. Thus the delocalization of the positive polaron is just one factor contributing to the efficiency of the cells.

An important question is whether our experimental data allows us to distinguish between delocalization of the positive polaron along the polymer chain (intrachain polaron) or also between polymer chains (interchain polaron), \textit{e.g.} if polymer chains stack. The X-ray crystal structure of P3HT suggests the possibility of the formation of \mbox{$\pi$-stacked} chains.\cite{Prosa:1992vy,Tashiro:1991bw,Brinkmann:2011gl,McCullough:1993dx} In films of P3HT, these spatial arrangements might be present. In order to probe how magnetic parameters will change upon P$^+$ delocalization between chains, we tested the simple model in which two trimers were constrained together similar to the X-ray crystal structure. As shown in Figure~S5, spin density calculations were performed on a dimer of stacked trimeric oligomers with the sulfurs constrained to remain a fixed distance apart. The system had an overall +1 charge to mimic the experimental conditions in concentrated solutions or films. For all distances (3.8--3.95~\AA), the calculated spin density is clearly delocalized over the entire complex with no preference for one of the two trimers. The DFT calculated principal values of the $g$-tensor in these dimers are similar to those of short oligomers (Supporting Information). This is expected as $g$-tensor axes of neighboring chains are parallel to each other, if P3HT molecules are highly ordered as found in crystalline phase. In the simplest model, the hf-interaction depends only on the number of thiophene units carrying the positive polaron; it does not matter if these thiophene units are part of the same chain or neighboring ones. In solution, the probability of the crystal-type stacking of polymer chains is lower, and interchain delocalization would be greatly diminished. However, the ENDOR spectra of P$^+$ are exactly the same as in the film. This suggests that the polaron is preferably delocalized along the chain, both in solution and in film. 

In conclusion, we investigated the electronic properties of light-induced positive and negative polarons formed in the PV-active polymer-fullerene blends. These included the ``reference'' polymer P3HT and representatives of two series of high-performance, low band gap polymers, PCDTBT and PTB7. We produced solar cells in our laboratory using the same polymers and fullerene derivatives (C$_{60}$-PCBM and C$_{70}$-PCBM) as in the EPR study. The solar cells obtained showed power conversion efficiencies of 3.3\,\% (P3HT), 6.1\,\% (PCDTBT) and 7.3\,\% (PTB7), which are comparable to the highest reported power conversion efficiencies of these high-performance polymers.\cite{nrel-gov,Park:2009kb,Liang:2010jb} Under illumination of these composites, two paramagnetic species are formed due to photo-induced electron transfer between the conjugated polymer and the fullerene. They are the positive, P$^+$, and negative, P$^-$, polarons on the polymer backbone and fullerene cage, respectively. To our knowledge, this work represents the first comprehensive EPR study of these low band gap polymers. High-frequency EPR (130 GHz) spectroscopy allowed resolving the EPR spectra of the individual paramagnetic species and to determine the principal components of the electronic $g$-tensors of positive and negative polarons. The $^1$H hyperfine coupling constants of these polarons were measured using the high spectral resolution of light-induced pulsed ENDOR spectroscopy. The experimental data obtained for the different polymers-fullerene composites provide an excellent reference database for validation of our and future theoretical calculations, and as a result, for elucidation of the electronic structures of the respective polarons. DFT calculations alone cannot reliably determine the electronic structure and delocalization of the positive polaron on the polymer for a variety of reasons, the most crucial ones mentioned below. Note, that the following reasoning is valid in general for DFT treatment of similar systems. First, the large size of calculated oligomers puts certain constraints on computational accuracy, in particular on the size of basis sets used. Second, the approximations inherent in the DFT method itself add a further uncertainty, which could lead to smaller or larger delocalization. Third, and very important, is that our DFT calculations were done in vacuo and do not include explicitly the interactions with nearest surrounding. In real systems, like films or frozen solutions of polymer:fullerene blends, interactions with surrounding molecules might substantially influence the conformations of the polymer, as well as delocalization of the charges along the polymer. While these factors prohibit to uncritically relying on the outcome of calculations, an estimation of the delocalization can be achieved by comparison of the calculated magnetic parameters for the various cases with the experimental values. The DFT method used here has shown in the past to provide magnetic parameters in good agreement for similar, but smaller system. We thus expect that the delocalization length of the positive polaron can be nicely estimated by comparison with the experimental data. 

In all cases, the calculated $g$-values showed no significant dependence on the length of the oligomer model when going beyond unrealistic small systems (like a P3HT monomer). The small difference between experimental and calculated $g$-values for larger, more realistic oligomer models is attributed to the three points mentioned above. Thus, while the agreement between experimental and calculated $g$-values is within the typical error range of DFT calculations for extended organic radical systems, the $g$-values themselves do not appear to be good parameters to determine the amount of polaron delocalization. On the contrary, the hyperfine interactions with protons show a clear dependence on the length of the oligomer model used in the calculation. The comparison with the experimental values revealed that in all three systems the positive polaron is distributed over rather large distances of 40--60~\AA ~on the polymer chain. This corresponds to about 15 thiophene units for P3HT, approximately three units PCDTBT, and about three to four units for PTB7. Previous EPR investigations of P3HT yielded a more localized polaron, which we attribute to the use of iodine as oxidant instead of light-induced polaron generation, which is identical with the processes in active OPV materials. No delocalization of the electron density over more than one fullerene molecule was observed experimentally. Analysis of the polaronic states in steady state condition do not reveal any significant broadening of the EPR line due to dipole--dipole or exchange interactions between unpaired electrons, which lead us to the conclusion that positive and negative charges under the study are separated for at least 25~\AA. Efficient delocalization of the positive polaron on the polymer seems to be the major reason for the efficient charge separation in bulk heterojunction systems as it minimizes charge recombination. It is interesting to note, that in natural photosynthetic systems a similar mechanism is invoked to optimize the yield of photoinduced electron transfer. Upon initial charge separation, the positive charge on the electron donor is delocalized over at least two large chlorophyll molecules to reduce the Coulomb interaction. This suppresses fast recombination and facilitates further transport of the electron, yielding long-lived charge separated states and resulting in the photosynthetic analogue of photocurrent. This work clearly demonstrates that advanced EPR spectroscopy in combination with modern concepts in DFT is an extremely powerful approach for investigation of light-induced charge dynamics in organic photovoltaic materials.

\section*{Abbreviations}

\begin{table}[htdp]
\begin{tabular}{p{.09\textwidth}p{.4\textwidth}}
CW      & continous wave \\
EPR     & electron paramagnetic resonance \\
ENDOR & electron nuclear double resonance \\
ET          & electron transfer \\
CT       & charge transfer \\
hf          & hyperfine \\
DFT      & density functional theory \\
MW      & microwave \\
RF       & radiofrequency \\
HF       & high frequency \\
OPV    & organic photovoltaic \\
P3HT   & poly(3-hexylthiophene-2,5-diyl) \\
PCDTBT   &   poly[[9-(1-octylnonyl)-9H-carbazole-2,7-diyl]-2,5-thiophenediyl-2,1,3-benzothiadiazole-4,7-diyl-2,5-thiophenediyl] \\ 
PTB7     &  poly[[4,8-bis[(2-ethylhexyl)oxy]benzo[1,2-b:4,5-b']dithiophene-2,6-diyl][3-fluoro-2-[(2-ethylhexyl)carbonyl]thieno[3,4-b]thiophenediyl]] \\ 
C$_{60}$-PCBM   &   [6,6]-phenyl-C61-butyric acid methyl ester \\
C$_{70}$-PCBM   &   [6,6]-phenyl-C71-butyric acid methyl ester \\
\end{tabular}
\end{table}%

\section*{Author Contributions}
The manuscript was written through contributions of all authors. All authors have given approval to the final version of the manuscript.

\section*{Acknowledgments}
 This material is based upon work supported as part of the Argonne-Northwestern Solar Energy Research (ANSER) Center, an Energy Frontier Research Center funded by the U.S. Department of Energy, Office of Science, Office of Basic Energy Sciences under Award Number DE-SC0001059 (JN, LU, and OGP) and by the DFG SPP ``Elementary processes in organic photovoltaics'', under contract DY18/6-1 (AS and VD). VD acknowledges financial support from ANSER during his research visit at ANL in 2012. The synthesis of PCDTBT (M.L. and S.B.) was supported by the Natural Sciences and Engineering Research Council (NSERC) of Canada. KLM was supported by the Department of Defense Army Research Lab (W911NF-0820039) and the National Institutes of Health (Grant 1SC2GM083717). BPB was supported by the National Science Foundation IL-LSAMP (Grant HRD-0413000) and GMG by the NIH/NIGMS (R25 GM059218). We thank Alexander F\"ortig and Carsten Deibel (both U of W\"urzburg) for solar cells fabrication and evaluation as well as for fruitful discussions.

\section*{Supporting Information}
Characteristic parameters of the polymer:fullerene BHJ solar cells; additional EPR spectra \& simulations; structures, energies, and magnetic parameters as obtained by DFT calculations. This material is available free of charge via the Internet at \href{http://www.rsc.org/suppdata/cp/c3/c3cp51477c/c3cp51477c.pdf}{www.rsc.org/pccp}. 

\bibliography{hex}

\end{document}